\documentclass[fleqn,usenatbib]{mnras}

\usepackage{newtxtext,newtxmath}

\usepackage[T1]{fontenc}

\DeclareRobustCommand{\VAN}[3]{#2}
\let\VANthebibliography\thebibliography
\def\thebibliography{\DeclareRobustCommand{\VAN}[3]{##3}\VANthebibliography}

\usepackage{mathtools,lipsum}

\usepackage{cuted}
\usepackage{float}
\usepackage{graphicx}	
\usepackage{amsmath}	
\usepackage[dvipsnames]{xcolor}

\title[Alignments around voids]{Intrinsic alignments of galaxies around cosmic voids}

\author[W. d'Assignies et al.]{
William d'Assignies D.$^{1}$\thanks{E-mail: wdoumerg@ens.fr}, Nora Elisa Chisari$^2$, Nico Hamaus$^3$, Sukhdeep Singh$^{4}$.
\\
$^{1}$Physics institute of the Ecole Normale Supérieure PSL, 45 rue d'Ulm, Paris, France.\\
$^{2}$Institute for Theoretical Physics, Utrecht University, Princetonplein 5, 3584 CC, The Netherlands.
\\
$^{3}$Universit\"ats-Sternwarte M\"unchen, Fakult\"at f\"ur Physik, Ludwig-Maximilians Universit\"at, Scheinerstr.~1, 81679 M\"unchen, Germany. \\
$^{4}$ McWilliams Center for Cosmology, Department of Physics, Carnegie Mellon University, Pittsburgh, PA 15213, USA.
}

\date{Accepted XXX. Received YYY; in original form ZZZ}

\pubyear{2021}

\begin{document}
\label{firstpage}
\pagerange{\pageref{firstpage}--\pageref{lastpage}}
\maketitle

\begin{abstract}
The intrinsic alignments of galaxies, i.e., the correlation between galaxy shapes and their environment, are a major source of contamination for weak gravitational lensing surveys. Most studies of intrinsic alignments have so far focused on measuring and modelling the correlations of luminous red galaxies with galaxy positions or the filaments of the cosmic web. In this work, we investigate alignments around cosmic voids. We measure the intrinsic alignments of luminous red galaxies detected by the Sloan Digital Sky Survey around a sample of voids constructed from those same tracers and with radii in the ranges: $[20-30; 30-40; 40-50]$ $h^{-1}$ Mpc and in the redshift range $z=0.4-0.8$. We present fits to the measurements based on a linear model at large scales, and on a new model based on the void density profile inside the void and in its neighbourhood. We constrain the free scaling amplitude of our model at small scales, finding no significant alignment at $1\sigma$ for either sample. We observe a deviation from the null hypothesis, at large scales, of 2$\sigma$ for voids with radii between 20 and 30 $h^{-1}$ Mpc, and  1.5 $\sigma$ for voids with radii between 30 and 40 $h^{-1}$ Mpc and constrain the amplitude of the model on these scales. We find no significant deviation at 1$\sigma$ for larger voids. Our work is a first attempt at detecting intrinsic alignments of galaxy shapes around voids and provides a useful framework for their mitigation in future void lensing studies. 
\end{abstract}
 



\section{Introduction}

Galaxies are known to align their shapes towards each other in the Universe \citep{Brown_2002,Mandelbaum_2006, hirata_red_blue,Blazek_2011,Singh,jacknife,samuroff2020advances,Pedersen_2020,Singh2021FP}. They also show preferential alignments with respect to the ``cosmic web'', the network of nodes and filaments that constitutes the structure of the Universe \citep{cosmic_web_yen_chi, Georgiou}. These alignments are most often present for luminous red galaxies and they constitute a regular contaminant to weak gravitational lensing observables in photometric samples \citep{Hirata&uros,krause}. Blue galaxies, on the contrary, show no significant shape alignment so far \citep{hirata_red_blue,mandelbaum_blue,10.1093/mnras/stz2197}, though numerical simulations suggest these could be detected by future surveys \citep{chisari_blue}. 

Models for intrinsic alignments rely on a connection between a galaxy shape and the tidal field \citep{Lamodel,Blazek_2011,EFT} at large scales. This connection is a linear dependency in the case of elliptical, pressure-supported, galaxies; and quadratic for spirals.\footnote{For simplicity and due to limitations in the data samples, elliptical galaxies are typically equated to red ones in the literature; and spirals are all assumed to be blue.} At small scales, alignments are modelled within the `halo model' framework \citep{Halo_model,Schneider_2010,fortuna2020halo}. This assumes that the distribution of satellite galaxies in a halo follows the spherically symmetric density profile of the dark matter, and that the orientations of their major axes are towards the centre of the halo.

Gravitational lensing refers to the deflections of photons as they travel through the gravitational potential of the large-scale structure of the Universe to our telescopes. ``Weak" gravitational lensing (corresponding to small and correlated distortions at approximately per-cent level) is one of the most promising observational techniques in order to elucidate the nature of dark matter and dark energy because it establishes a relation between apparent shapes of observed objects and the matter density field throughout the history of the Universe (\citealt{kaiserWLDM}). Therefore, several experiments have made weak lensing a key part of their programs. Currently ongoing ones are the Kilo-Degree Survey \citep{KIDS}, Hyper Suprime-Cam
\citep{Aihara_2017} and the Dark Energy Survey
\citep{collaboration2020dark}. Planned to start early in this decade are the Vera Rubin Observatory  \citep{LSST} and {\it Euclid} \citep{EUCLID}. 

Although most studies of weak gravitational lensing focus on the effect of matter overdensities in background galaxy shapes, the lensing by voids is emerging as a useful tool to constrain cosmology \citep{NH1,NH2,NH3,NH4,lensingvoids,NH5,WLvoids,NH6}. Several studies \citep{Barreira_2015,wlvoidtest} suggested that lensing by voids could be used as a test of general relativity. Voids offer other possibilities for testing the cosmological model, including the presence of massive neutrinos \citep{NH7,NH8,NH9,void_neutrino}, probing gravity through redshift-space distortions \citep{void_gravity,NH10,Hamaus16,NH11,NH12,NH13}, studying the nature of dark matter \citep{NH14,NH15,NH16,pisani2019cosmic} and looking for signatures of primordial non-Gaussianity \citep{NH17}. In the context of photometric galaxy samples, it is yet unclear whether and how galaxy alignments could contaminate these type of studies. 

On the other hand, intrinsic alignments have been proposed as a probe of cosmology in their own right. Examples of potential applications include probing inflationary models \citep{inflation,non-gaussianity,inflation2}, gravitational waves present in the early Universe \citep{chisaribicep,biagetti}, baryon acoustic oscillations \citep{BAOChisari}, the role played by neutrinos in the evolution of the Universe \citep{neutrinoIA}, or modifications of gravity \citep{reischke2021testing}. Overall, intrinsic alignments encode information in a complementary manner to other probes of the large-scale structure. 

\begin{figure}
    \centering
    \includegraphics[scale=0.6]{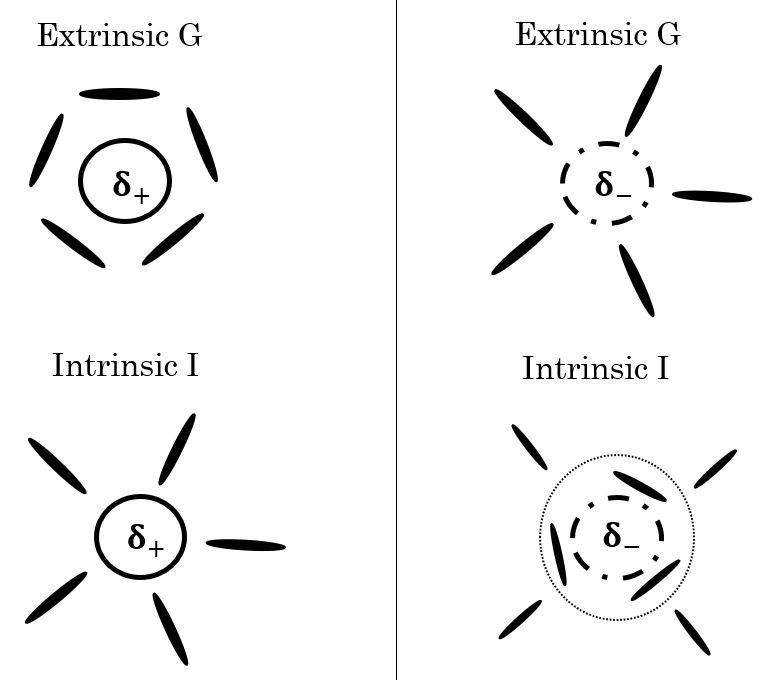}
    \caption{Galaxy shapes (ellipses) patterns for intrinsic (I) alignments and extrinsic gravitational (G) lensing. On the left column, we show the case of a massive structure at the center of the image:  an overdensity $\delta_+$. On the right column, we show the case of a cosmic void: an underdensity $\delta_-$. The bottom panels assume a positive void bias.}
    \label{fig:paternsGLIA}
\end{figure}

Motivated by significant detections of red galaxy alignments around overdensities in the matter field,we would similarly expect red galaxies to be aligned with cosmic voids. Such effects could potentially act as a contaminant to void lensing studies, or a cosmological probe. We will see shortly that our model predicts that galaxy shapes align around voids tangentially near the void, and radially (tangentially) far from the void, for a positive (negative) void bias, as illustrated in Figure \ref{fig:paternsGLIA}.
While we focus on the alignments of red galaxy shapes around voids, there is previous evidence that blue galaxies align their rotation axes preferentially with respect to voids. \citet{spiral_gal_CV1,spiral_gal_CV3} have reported $>95\%$ confidence level detections of this mode of alignment (although \citealt{spiral_gal_CV2} have found this to be consistent with null). This type of alignment could, as well, constitute an eventual contaminant to void lensing but are not considered here.

In this work, we take the first steps towards the modelling and measurement of the alignments of the shapes of galaxies around voids using a sample of 3192 voids with redshifts $0.4<z<0.8$, and effective radii between 20 and 50 $h^{-1}$ Mpc. The study of alignments around cosmic voids is driven by two motivations. First, we expect that the assumption that galaxy shapes respond linearly to the tidal field should be very accurate. In other words, voids would offer an excellent environment where to test the linear model of galaxy alignments. This is motivated by the findings of \citet{voidbias}, who demonstrated that the clustering of tracers around voids is very accurately described by a linear model, in contrast to the auto-correlation of the positions of those tracers. Second, we envisage that alignments of galaxies around voids could help constrain void properties such as their density profile, or their bias, and help enable some of their cosmological applications.  \citealt{environmental_dependance} already investigate the environmental dependence of the intrinsic ellipticity of spiral and elliptical galaxies, respectively, with a Gaussian random density field. The dependence on environment is modelled by the number of positive eigenvalues of the tidal tensor, which allows a differentiation between voids, sheets, filaments and superclusters. They find that alignment around voids is weak - an order of magnitude lower than for other environments. In this work, we develop our own theoretical model for alignment around voids, and contrast this against observations. 

This paper is organized as follows. In section \ref{sec:2}, we describe the linear alignment model  along with our void-matter power-spectrum model. Section \ref{sec:3} describes the galaxy and void catalogues, and the estimator we use to measure the two-point correlation function. We present our results in section \ref{sec:4}, analyze them in section \ref{sec:discussion} and conclude in section \ref{sec:5}. For all of the model predictions, we use the following cosmology: $h=0.7$, $\Omega_{\text{CDM}}=0.225$,  $\Omega_{\text{b}}=0.045$, $\Omega_{\Lambda}=0.73$, $\sigma_8=0.8$, $n_s=1$. We measure distances in $h^{-1}$ Mpc. This work used version 2.1 of the CCL library \citep{Chisari2019} to compute essential cosmological functions, including matter power spectra derived from the CAMB software \citep{camb}.

\section{Formalism}
\label{sec:2}

\subsection{Weak gravitational lensing}

Working under the flat-sky approximation, we introduce a Cartesian frame defined by unit vectors $\hat{\bf e}_x$ and $\hat{\bf e}_y$ in the projected plane and $\hat{\bf e}_z$ along the line-of-sight. To describe the shape of a galaxy, we introduce two quantities: $\gamma_1$ and $\gamma_2$. $\gamma_2$ represents the stretching along $\hat{\bf e}_x$ and $\gamma_1$ along the axis $\hat{\bf e}_x$ rotated by $-45$ degrees.

In order to describe the shape relative to the separation vector between two points, corresponding to the coordinates of a galaxy and a void, we introduce the angle $\phi$ between this line and the $\hat{\bf e}_x$ axis. The shape in this frame is described by the rotation of $\gamma=(\gamma_1,\gamma_2)$ by an angle $2\phi$, as
\begin{equation}
     \begin{split}
         \gamma_+&=-\cos(2\phi) \gamma_1 -\sin(2 \phi)\gamma_2,\\
         \gamma_\times&=\sin(2 \phi)\gamma_1 -\cos(2 \phi)\gamma_2.
     \end{split}
 \end{equation}
Thus, a positive $\gamma_+$ corresponds to a radial alignment.

In practice, the ellipticity of a galaxy is measured as 
\begin{equation}
    (\epsilon_+,\epsilon_\times)=\frac{1-(b/a)^2}{1+(b/a)^2}(\cos(2\phi),\sin(2\phi)).
\end{equation}
where $b$ is the length of the minor axis and $a$, the length of the major axis. 
For a given galaxy $i$, the observed shape can be decomposed in the initial ellipticity $\epsilon_{\text{source}}^i$, assumed to be random, an intrinsic ellipticity, $\gamma^I_i$, distorted by the large-scale structure, and the distortion due to weak lensing, $\gamma^G_i$ (see \citealt{PhysRevD.70.063526})\footnote{Formally, one needs to take into account a responsivity factor in connecting measured ellipticities to the lensing shear \citep{BJ02}. In this work, because we are only interested in assessing a potential detection of intrinsic alignments, this factor is not included in the calculations.}:
\begin{equation}
    \epsilon_{\text{obs}}^i=\epsilon_{\text{source}}^i+\gamma^I_i+\gamma^G_i.
\end{equation}
$\gamma^I_i$ and $\gamma^G_i$ are correlated with the overdensity of the matter field, $\delta$. Thus, taking the average over all observed galaxies,
\begin{equation}\label{eq:contam}
    \langle \delta \epsilon^j_{\text{obs}} \rangle=\langle \delta~ \gamma^I_j\rangle+\langle \delta~ \gamma^G_j\rangle.
\end{equation}
Eq. (\ref{eq:contam}) illustrates how intrinsic alignments act as a weak lensing contaminant. 

\subsection{Voids}
\label{ss:voids}
The void density profile is well-described by the function \citep{Hamaus_2014} 
\begin{equation}\label{eq:vd}
    \Delta_\mathrm{v}(r)\equiv \frac{\rho_\mathrm{v}(r)}{\bar{\rho}_\text{m}}-1=\delta_c\frac{1-\left(r/r_s\right)^\alpha}{1+\left(r/r_\mathrm{v}\right)^\beta} ,
\end{equation}
with 4 free parameters $\delta_c$, $r_s$, $\alpha$, $\beta$, and the measured void radius $r_\mathrm{v}$. $\delta_c<0$ is the central density contrast of the void, $\rho_\mathrm{v}$ and $\bar{\rho}_\text{m}$ are the matter density field around the void centre and its average background value, respectively. Spherical symmetry is warranted when one averages over a sufficiently large number of  arbitrarily shaped voids with random orientations.
For $r> 2 r_\mathrm{v}$, $\rho_\mathrm{v}(r)\approx\bar{\rho}_\mathrm{m}$, and the average void density profile quickly converges to the average density of the Universe. At the void centre, $\rho_\mathrm{v}(0)=\bar{\rho}_\mathrm{m}(1+\delta_c)$, which represents a drop in density with respect to the average.   
$\Delta_\mathrm{v}$ is negative at the centre of the void but can change sign near $r\sim r_\mathrm{v}$ \citep{Hamaus_2014}. Far from the centre of the void, $\Delta_\mathrm{v}(r)\sim 0$ due to the convergence to  the mean density of the Universe. 
Typical values of the free parameters in Eq. (\ref{eq:vd}), derived from simulated mock catalogues, are: $( r_s/r_\mathrm{v}, \alpha, \beta, \delta_c)=(0.82, 1.6, 9.1, -0.36)$  \citep{hamaus2020precision}. These parameters correspond to a population of mock galaxies with bias of $b_\mathrm{g}\approx 2.2$. For our sample (Table \ref{tab:data}), we expect a slightly lower bias, about 1.85, and thus the value of $\delta_c$ could be slightly different than the one we assume.

\subsection{Linear alignment model}
\label{sec:gamma_formalism_LA} 

\citet{Lamodel} proposed a linear relation between the shape of a galaxy and the tidal field of the Universe, expressed as 
\begin{equation}
   \gamma_{+,\times}({\bf x})= -\frac{A_IC_1}{4\pi G}\left(\nabla_x^2-\nabla_y^2~,~2\nabla_x\nabla_y\right)S\left[\Psi\right],
\end{equation}
where $A_IC_1$ is a constant representing the response of a galaxy shape to a tidal deformation, and $S$ is a filter which cuts off fluctuations of the gravitational potential $\Psi$ at small scales. $C_1$ is retained for historical reasons: it is the value of  the first intrinsic alignment measurement by \citet{Brown_2002}. Equivalently, in Fourier space,
\begin{equation}
\begin{split}
    \gamma_{+,\times}({\bf k})&=\frac{A_IC_1}{4\pi G}(k_x^2-k_y^2,2k_xk_y)\Psi\\\label{eq:expre}
    \end{split}
\end{equation}
where $\Psi$ is the Fourier transform of the gravitational potential,
\begin{equation}
\begin{split}
    \Psi({\bf k})&=-4\pi G \frac{\bar{\rho}^0_\text{m}}{D(z)}\frac{\delta_{\text{m}}({\bf k})}{k^2}
\end{split}
\end{equation}
 with $\bar{\rho}^0_\text{m}$, the mean density of the Universe at redshift $z=0$, and $D(z)$, the growth function (normalized to unity at $z=0$).
The alignment bias can depend on many parameters such as the dynamical properties of a galaxy, its merger history, its redshift or its environment. 
Because this model is linear, it is only valid for large scales where the density fluctuations are small.  We will neglect the filter $S$ in the rest of this work, because we will mainly focus on modelling linear scales where it does not have a significant impact.

\subsection{Correlation functions}

 We introduce $\delta_\text{m}$, the matter overdensity, and $\delta_\mathrm{v}$, the void distribution, 
\begin{align}
 \delta_\text{m}({\bf x})&=\left(\rho_\text{m}({\bf x})-\bar{\rho}_\text{m}\right)/\bar{\rho}_\text{m}\label{eq:deltam}\\
 \delta_\mathrm{v}({\bf x})&=\frac{1}{N_\mathrm{v}^\mathrm{tot}}\sum_{i\in V}\delta^{(3)}(\bf x-\bf x_i)\label{eq:deltav}
\end{align}
where  $V$ is the void sample, $\delta^{(3)}$ is the Dirac function, $N_\mathrm{v}^\mathrm{tot}$ the number of voids and $\{\bf x_i\}$ are the void centres.  The void-matter power spectrum is defined as the correlation of $\delta_\text{m}$ and $\delta_\mathrm{v}$ \citep{Hamaus_2013}, 
\begin{equation}
\langle \delta_\mathrm{v} ( {\bf k} ) \delta_{\text{m}}({\bf k'})\rangle  = \delta^{(3)} ({\bf k}-{\bf k'} )   P_{\mathrm{vm}}({\bf k} )(2\pi)^3.
\end{equation}

The correlation functions between galaxy shapes and the void distribution are
\begin{align}
    \xi_{\mathrm{v}+}({\bf r})&=\langle \delta_\mathrm{v}({\bf x}) \gamma_+({\bf x}+{\bf r})\rangle\\
    \xi_{\mathrm{v}\times}({\bf r})&=\langle \delta_\mathrm{v}({\bf x}) \gamma_\times({\bf x}+{\bf r})\rangle,
\end{align}
where $\langle \ldots \rangle$ represents an average over many realizations of the Universe. The void-intrinsic galaxy shape correlation is given by
\begin{equation}
\label{eq:corr1}
    \begin{split}
        \xi_{\mathrm{v}+}({\bf r})&=-\frac{A_IC_1\rho_c  \Omega_\text{m}}{D(z)}\\
        &\times \iint \frac{d^3kd^3k'}{(2\pi)^6}\langle \delta_\mathrm{v}({\bf k}) \delta_\text{m}({\bf k}')\rangle \frac{k_{x}^2-k_{y}^2}{k^2}e^{i(k_xr_p+k_z\Pi)}.
    \end{split}
\end{equation}
with $\Omega_\text{m}=\bar{\rho}^0_\text{m}/\rho_c$, $\rho_c$ being the critical density and  ${\bf r}=r_p\hat{\bf e}_x+\Pi\hat{\bf e}_z$, with $r_p$ the component perpendicular to the line-of-sight,  and $\Pi$ the component along the line-of-sight. The x-axis has been defined as the separation axis on the sky.  \footnote{In reality, what we measure is not $\gamma_I$ directly, but sampled at the locations of galaxies. Hence, the actual observable is  $\tilde{\gamma}_I=(1+\delta_\mathrm{g})\gamma_I$ \citep{Blazek_2011}. Since $\delta_\mathrm{g}$ is proportional to $\delta_\text{m}$, this term is higher than linear order and we will neglect it here. Further details can be found in Appendix \ref{sec:appendixA}. Here the $x$-axis has been defined to coincide with the direction of the separation vector projected on the sky.}

We can simplify Eq. (\ref{eq:corr1}) by integrating over the angle in $\vec k_\perp$ space, with $k_\perp=\sqrt{k_x^2+k_y^2}$. Using $k_x=k_\perp\cos(\theta) $ and $k_x^2-k_y^2=k_\perp^2\cos{(2\theta)}$, we recognize the second Bessel function $J_2$, and obtain the expression
\begin{equation}
\begin{split}
\xi_{\mathrm{v}+}({\bf r})&=\frac{A_IC_1\rho_c  \Omega_\text{m}}{\pi^2D(z)} \\
&\times\int_0^{\infty}\int_0^{\infty} dk_\perp dk_z P_{\mathrm{vm}}(k)\frac{k_\perp^3}{k^2}\cos(k_z\Pi )J_2(k_\perp r_p).\label{eq:xivp}
\end{split}
\end{equation}
Similarly, one can show that $\xi_{\mathrm{v}\times}=0$.
Observational studies most often measure the projected correlation function, obtained by integrating Eq. (\ref{eq:xivp}) over the line-of-sight,
\begin{equation}
    \omega_X(r_p)=\int_{-\Pi_{\text{max}}}^{+\Pi_{\text{max}}}\xi_X(r_p,\Pi)d\Pi,
\end{equation}
with $X\in \{\mathrm{v}+,~\mathrm{v}\times \}$. We chose $\Pi_\text{max}$=150 $h^{-1}$ Mpc to ensure that we capture the alignments in the environments of voids.\footnote{For groups or galaxy alignments, the literature most often quotes values of $60 h^{-1}$ Mpc $<\Pi_\text{max}<100 h^{-1}$ Mpc.}  Obviously, $\omega_{\mathrm{v}\times}=0$, and 
\begin{equation}
\begin{split}
\omega_{\mathrm{v}+}(r_p)=&\frac{A_I}{\pi^2}\frac{C_1\rho_c \Omega_\text{m}}{D(z)}\\
&\times\iint_0^\infty  dk_\perp dk_z \frac{k_\perp^3}{k^2 k_z}P_{\mathrm{vm}}(k)\sin(k_z\Pi_\text{max})J_2(k_\perp r_p).\label{eq:wproj1}
\end{split}
\end{equation}

\subsection{Window function and redshift-space distortions}

Eq. (\ref{eq:wproj1}) for $\omega_{\mathrm{v}+}$ was implicitly given at a specific redshift. We introduce $p_\mathrm{v}(z)\equiv 1/N_\mathrm{v}^{\text{tot}} dN_\mathrm{v}/dz$ and $p_\mathrm{g}(z)\equiv 1/N_\mathrm{g}^{\text{tot}} dN_\mathrm{g}/dz$, the redshift distributions of voids and galaxies, respectively, and the corresponding redshift window function \citep{mandelbaum_blue}:
\begin{equation}
    W(z)=\frac{p_\mathrm{v}(z)p_\mathrm{g}(z)}{\chi^2d\chi/dz}\left(\int \frac{p_\mathrm{v}(z)p_\mathrm{g}(z)}{\chi^2d\chi/dz}dz\right)^{-1}.
\end{equation}
We integrate Eq. (\ref{eq:wproj1}) to yield the final prediction for the observable,
\begin{equation}
\begin{split}
    \omega_{\mathrm{v}+}(r_p)=&\frac{A_IC_1\rho_c \Omega_\text{m}}{\pi^2}\int dz \frac{W(z)}{D(z)}\int_0^{+\infty}dk_\perp \int_0^{+\infty}dk_z\\
     &\left\{\frac{k_\perp^3}{k^2 k_z}P_{\mathrm{vm}}(k,z)\sin(k_z\Pi_\text{max})J_2(k_\perp r_p)\right\}.\label{eq:wproj2}
    \end{split}
\end{equation}
One might consider a correction to Eq. (\ref{eq:wproj2}) due to the presence of redshift-space distortions, first derived by \cite{kaiser_effect} and consisting of a Doppler shift due to peculiar velocities. Including redshift-space distortions in our analysis would yield a correction smaller than 10 $\%$ at very large scales ($r_p \sim 100$ $h^{-1}$ Mpc). Since this correction is significantly smaller than the size of our error bars, we neglect it. 
Finally, as we restrict this study to galaxies in the immediate vicinity of voids ($<\,150\,h^{-1}$ Mpc), the lensing contribution is expected to be small.
\subsection{Void-matter power spectrum}

To determine the void-matter power spectrum, we will consider two regimes. The intermediate-scale regime will be used for scales $r_p\sim r_\mathrm{v}$, i.e. near the void boundary. A second regime will describe large scales, i.e. $r_p$ far enough from the void (typically $r_p> 2r_\mathrm{v}$). We do not consider a small-scale regime $r_p\ll r_\mathrm{v}$, since by definition this region contains very few galaxies resulting in a large statistical uncertainty of our measurement. The combination of both models gives the alignment pattern of bottom-right panel of Figure \ref{fig:paternsGLIA} (for a positive void bias).

\subsubsection{Intermediate-scale regime}
\label{sec:model_intscale}

For a single void of radius $r_\mathrm{v}$ centred at the origin, if $r\sim r_\mathrm{v}$, we can consider that the matter distribution is well-described by the void density profile, $\rho_\mathrm{v}(r)$. In this case, the void-matter two-point correlation is given by \citep{void_gravity,voidbias}
\begin{equation}
   \langle \delta_\mathrm{v}({\bf x}-{\bf 0}) \delta_\mathrm{m} ({\bf x}+{\bf r})\rangle= \frac{\rho_\mathrm{v}(r)}{\bar{\rho}_\text{m}}-1=\Delta_\mathrm{v}(r).
\end{equation}
Thus, the power spectrum between void centres and the matter distribution at intermediate scales is \citep{Chan_2014}
\begin{equation}
   \Delta_\mathrm{v}(k)\equiv P^{\mathrm{is}}_{\mathrm{vm}}(k)=\int_0^{+\infty}4\pi r^2dr\frac{\sin(kr)}{kr}\Delta_\mathrm{v}(r)\label{eq:dv}
\end{equation}
An empirical expression of $\Delta_\mathrm{v}$ in real space is obtained from simulations and only describes scales smaller and equivalent to the size of the void. 
To ensure the consistency of our separation of scales, we checked numerically that the contribution of the usual power spectrum with a cutoff at $k>\pi/2r_v$ (a basic model for large scale contamination) is negligible. Thus there is no large-scale contribution to Eq. (\ref{eq:wproj2}) at $r_p\sim r_\mathrm{v}$. The mathematical reason for the large-scale contribution to be suppressed is that $J_2(k_\perp r_p)$ kernel peaks near $k_\perp r_p\sim \pi$. 
This case is similar to the usual Limber approximation.

The power spectrum depends on the void radius, which motivates us to split our data into narrow radius intervals of 10 $h^{-1}$ Mpc. 
To account for potential mismatches between $\Delta_\mathrm{v}$ as predicted by simulations, and our observations, e.g. coming from different values of the density contrast of the voids, as detailed in Section \ref{ss:voids}, we introduce a free parameter $a_\mathrm{v}$ by assuming $\Delta_\mathrm{v}^{\text{data}}(k)=a_\mathrm{v}\Delta_\mathrm{v}^{\text{sim}}(k)$. 
For intermediate scales, we obtain the following model for the void-intrinsic shape correlation,
\begin{equation}
    \begin{split}
    \omega_{\mathrm{v}+}(r_p)=&A_I^Va_\mathrm{v}\frac{C_1\rho_c \Omega_\text{m}}{\pi^2}\int dz \frac{W(z)}{D(z)}\int_0^{+\infty}dk_\perp \int_0^{+\infty}dk_z\\
     &\left\{\frac{k_\perp^3}{k^2 k_z}\Delta_\mathrm{v}^{\text{sim}}(k)\sin(k_z\Pi_\text{max})J_2(k_\perp r_p)\right\},
    \end{split}
    \label{eq:int_scale_pred}
\end{equation}
with $A_I^V$ the alignment coefficient corresponding to void neighbourhood\footnote{As we mention in section \ref{sec:gamma_formalism_LA}, the alignment coefficient depends on many parameters including the environment.}. Notice we have not considered any explicit redshift dependence in the parameters that characterize the voids, and we instead hold them constant over our redshift range of interest, $0.4<z<0.8$.

\subsubsection{Large-scale regime}
\label{sec:model_largescale}

The void-matter power spectrum can be split into two terms \citep{Chan_2014},
\begin{equation}
\begin{split}
    P_{\mathrm{vm}}(k)=&\int_0^{+\infty}dr\,4\pi r^2\frac{\sin(kr)}{kr}\left[\Delta_\mathrm{v}(r)-b_\mathrm{v}(r_\mathrm{v})\xi_{\mathrm{mm}}^{\text{lin}}(r)\right] \\
    &+b_\mathrm{v}(r_\mathrm{v})\int_0^{+\infty}dr\,4\pi r^2\frac{\sin(kr)}{kr}\xi_{\mathrm{mm}}^{\text{lin}}(r),
    \end{split}
    \label{eq:pvm}
\end{equation}
where $b_\mathrm{v}$ is the void bias \citep{Hamaus_2013}, and $\xi^{\text{lin}}_{\mathrm{mm}}$ is the matter correlation function predicted from linear theory. 

In the large-scale limit, the void-matter power spectrum should approach the shape of the matter-matter power spectrum. This implies there exists a radius $r_\star$ such that $\Delta_\mathrm{v}(r>r_\star)=b_\mathrm{v}(r_\mathrm{v})\xi_{\mathrm{mm}}^{\text{lin}}(r)$. Therefore, Eq. (\ref{eq:pvm}) can be simplified to 
\begin{equation}
    \begin{split}
    P^\mathrm{ls}_{\mathrm{vm}}(k)=&\int_0^{r_\star}dr\,4\pi r^2\frac{\sin(kr)}{kr}\left[\Delta_\mathrm{v}(r)-b_\mathrm{v}(r_\mathrm{v})\xi_{\mathrm{mm}}^{\text{lin}}(r)\right]\\
    &+b_\mathrm{v}(r_\mathrm{v})P_{\mathrm{mm}}^{\text{lin}}(k).
    \end{split}
    \label{eq:pvm2}
\end{equation}
For small wave vectors (large scales), the first term can be neglected, and
we recover the familiar linear bias result,
\begin{equation}
    P^\mathrm{ls}_{\mathrm{vm}}\left(k< \frac{2\pi}{r_\mathrm{v}}~~\vert r_\mathrm{v},~z\right)=b_\mathrm{v}(r_\mathrm{v})P_{\mathrm{mm}}^{\text{lin}}(k,z).
\end{equation}
Here again, the power spectrum depends on void radius (as $b_\mathrm{v}$ does), which justifies our selection of  void radius intervals of 10 $h^{-1}$ Mpc for the large-scale measurements. The resulting projected correlation function in the large-scale regime is given by
\begin{equation}
    \begin{split}
    \omega_{\mathrm{v}+}(r_p)=&A_I^Vb_\mathrm{v}(r_\mathrm{v})\frac{C_1\rho_c \Omega_\text{m}}{\pi^2}\int dz \frac{W(z)}{D(z)}\int_0^{+\infty}dk_\perp \int_0^{+\infty}dk_z\\
     &\left[\frac{k_\perp^3}{k^2 k_z}P_{\mathrm{mm}}^{\text{lin}}(k,z)\sin(k_z\Pi_\text{max})J_2(k_\perp r_p)\right].
    \end{split}
    \label{eq:large_scale_pred}
\end{equation}


\section{Methodology}
\label{sec:3}

\subsection{Data sets}

We introduce the following notations for samples used in this work. $V$ represents the void sample positions and $R_V$, the corresponding randoms. $S_+$  represents the shapes and positions of the sample of galaxies, and $R_S$, the corresponding randoms. 

\subsubsection{Galaxy positions and shapes}
\label{sec:gxycat}
We use a sample of galaxies from data release 12 (DR12) of the  SDSS-III  Baryon Oscillation Spectroscopic Survey (BOSS, \citealt{SDSS3}). We consider the CMASS sample: massive galaxies at redshift $0.4<z<0.8$. We use the catalogue of galaxy shapes complied by \cite{Singh2021FP} by matching the CMASS sample with the galaxy shape measurements from \cite{Reyes2012}. We also make use of the random catalogue provided by the BOSS collaboration, which is a realization of an unclustered galaxy distribution with the same survey geometry (mask). We use 13 
times more random points than galaxies at any given redshift to avoid the size of the random catalogue having an impact on the observables. 

\subsubsection{Void positions}
The void catalogue is obtained using the {\sc vide} software\footnote{\url{https://bitbucket.org/cosmicvoids/vide_public/}}~\citep{VIDE} applied to the combined sample of DR12 BOSS galaxies previously used in \cite{hamaus2020precision}. {\sc vide} is based on {\sc zobov}~\citep{ZOBOV}, an algorithm that identifies voids by searching for ``basins'' in the density field of the observed tracers. Each location within the survey volume is attributed to its closest tracer (i.e., a galaxy). This operation defines a Voronoi tessellation with cells of volume $\mathcal{V}_i$ for each tracer $i$. The density field of tracers $\rho({\bf x})$ at any given location ${\bf x}$ inside the survey volume is then estimated by evaluating $1/\mathcal{V}_i$. {\sc vide} identifies voids as watershed basins of arbitrary shape, uncovering extended underdense regions that are surrounded by overdense boundaries. Every Voronoi cell contained in such a basin is then considered as part of a void.

\begin{figure}
    \centering
    \includegraphics[width=0.47\textwidth]{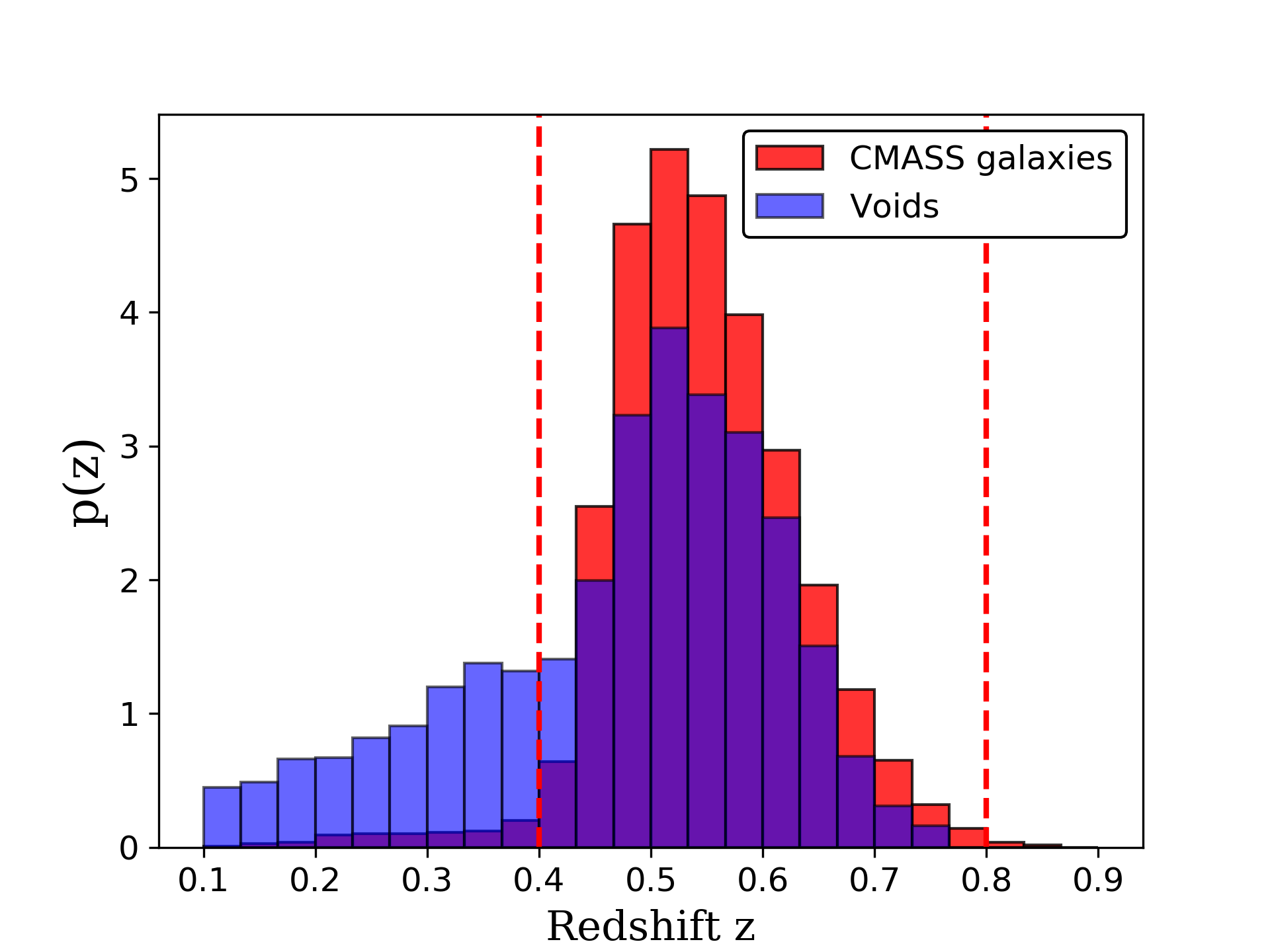}
    \caption{Redshift distribution of CMASS galaxies (red) and voids (blue). Red dashed lines represent the redshift range used in our study. The distribution is normalized such that $\int_z p(z)dz=1$. The presence of voids at $z<0.4$ is due to the catalogue being constructed from the full DR12 BOSS data. In practice, we are only interested in cross-correlations between physically close pairs of galaxies and voids.}
    \label{fig:cmass_z}
\end{figure}

The location of the void centre ${\bf X}$ is defined as the volume-weighted barycentre among all of its constituent Voronoi cells: ${\bf X}=\sum_i{\bf x}_i\mathcal{V}_i/\sum_i \mathcal{V}_i$, where ${\bf x}_i$ is the location of each tracer that defines the void. The effective void radius is defined as $r_\mathrm{v}=(3/4\pi\sum_i\mathcal{V}_i)^{1/3}$. The galaxy and void catalogues cover the same region of the sky and redshift range. In Figure \ref{fig:cmass_z}, we present their redshift distributions, which are very similar over the CMASS redshift range. Voids appear equally distributed within that galaxy sample.

Finally, we use a random catalogue for void positions, $R_V$, which has been generated with same redshift and void radii distributions, and takes into account the survey mask. This random catalogue is 10 times larger than the void one.

\subsection{Estimators}
We measure the correlation function of void positions and galaxy shapes using a Landy-Szalay estimator \citep{operatorIA},
\begin{equation}\label{eq:estimator}
    \xi_{\mathrm{v}+}(r)=\frac{VS_+-R_VS_+}{R_VR_S}
\end{equation}
where the terms $VS_+$, $R_VS_+$ and $R_VR_S$ stand for the sums\footnote{The condition $\vert i-j\vert=r_p,\Pi$ means that the vector between objects $i$ and $j$ is within a certain range of projected value $r_p$ and value $\Pi$ along the line-of-sight.} 
\begin{equation}
\begin{split}
    VS_+(r_p,\Pi)&=\sum_{\substack{i\in S,j \in V\\ \vert i-j \vert=r_p, \Pi }} \gamma_+^{(i)}\langle i\vert j\rangle\\
    R_VS_+(r_p,\Pi)&=\sum_{\substack{i\in S,j \in R_V\\ \vert i-j \vert=r_p, \Pi }} \gamma_+^{(i)}\langle i\vert j\rangle
    \end{split}
\end{equation}
where $\gamma_+^{(i)}\langle i\vert j\rangle$ denotes the $+$ shape of a galaxy $i$ relative to void $j$. $R_VR_S(r_p, \Pi)$ is the number of pairs of random voids and random galaxies that are separated by $r_p$ and $\Pi$. We numerically normalize the terms $R_VS_+$ and $R_VR_S$ to account for the relative size of random and galaxy catalogues. 

To obtain the projected correlation function, we divide the line-of-sight interval 
into $n$ bins of length $2\Delta\Pi^{(m)}$ and centred at  $\Pi^{(m)}$,
\begin{equation}\label{eq:estimfinal}
    \omega_{\mathrm{v}+}(r_p)\approx\sum_{m=1}^n \xi_{\mathrm{v}+}(r_p, \Pi^{(m)})2\Delta\Pi^{(m)}.
\end{equation}

The different samples used in the estimation of the correlation functions are described in Table~\ref{tab:data}. We divide the void sample according to their sizes in intervals of small voids: $[20,30]\,h^{-1}$ Mpc, intermediate voids: $[30,40]\,h^{-1}$ Mpc and large voids: $[40,50]\,h^{-1}$ Mpc. The choice of minimum radius is driven by the fact that smaller voids are surrounded by overdensities (e.g. clusters) which can induce a very strong correlation and wash out the void signal. We also had at our disposal voids with a radius larger than 50 $h^{-1}$ Mpc, but the number of these voids per interval of 10$ h^{-1}$ Mpc is significantly smaller. Since our measurements are already not significant for the voids with  radii in 40-50 $h^{-1}$Mpc range, we did not extend this study to these larger voids.

\begin{table*}
	\centering
	\begin{tabular}{lcccccr} 
		\hline
		Sample & Small voids & Intermediate voids  & Large voids  & Galaxies\\
		 &  $r_\mathrm{v}=[20,30]\,h^{-1}$ Mpc & $r_\mathrm{v}=[30,40]\,h^{-1}$ Mpc & $r_\mathrm{v}=[40,50]\,h^{-1}$ Mpc & \\
		\hline
		Data&1089&1220&883&560011\\
		Random & 11777 &12021 &8485& 7828387\\
		\hline
	\end{tabular}
	\caption{Characteristics of the various samples used in estimating the correlation function of galaxies around voids (Eq. \ref{eq:estimfinal}). All samples span the redshift range $0.4<z<0.8$.\label{tab:data}}
\end{table*}

\subsection{Covariance matrix}

 We use the jackknife method to obtain estimates of the covariance matrix and error bars for the projected correlation function of void positions and galaxy shapes. We split the observed area of the sky into $N_{\text{reg}}$ sub-regions taking the survey mask into account. We measure a set of projected correlation functions, $\tilde{\omega}_{\mathrm{v}+}^{i}(r_p)$, by removing one sub-region $i$ at a time. We define $\bar{\omega}_{\mathrm{v}+}(r_p)$ as the mean of $\tilde{\omega}_{\mathrm{v}+}^{i}(r_p)$. The covariance matrix $\mathcal{C}[r_p^i,r_p^j]$ is estimated via
 \begin{equation}
     \mathcal{C}[r_p^i,r_p^j]=\frac{N_{\text{reg}}-1}{N_{\text{reg}}}\sum_{ k=1 }^{N_{\text{reg}}}\left[\tilde{\omega}^k(r_p^i )-\bar{\omega}_{\mathrm{v}+}(r_p^i)\right]\left[\tilde{\omega}^k(r_p^j )-\bar{\omega}_{\mathrm{v}+}(r_p^j)\right].
 \end{equation}
The diagonals represent the error bars shown in the figures, $\sigma(r_p^i)=\sqrt{\mathcal{C}[r_p^i,r_p^i]}$.
These expressions provide accurate estimates of the uncertainty in the estimated projected correlation function, if sub-regions satisfy the following properties:
 \begin{enumerate}
     \item the sub-volume sizes are larger than the biggest scale probed in the observable,
     \item $N_{\text{reg}}$ is larger than the number of bins where the projected correlation function is estimated, and
     \item the number of data points (e.g. voids and galaxies) in each sub-volume is approximately the same.
 \end{enumerate}
As the number of voids is limited (a few thousand), in order for them to be distributed equally among the sub-regions, the number of sub-regions must be kept small. We use 64 sub-regions as a compromise to meet the above conditions. To determine the actual sub-regions, we use a $K$-means algorithm on the sphere to find maximally separated centres in a RA and DEC distribution of randoms $R_S$. The choice of $R_S$ (vs. $R_V$) for determining the jackknife sub-regions is due to higher sampling and justified by the fact that the voids are selected from the same galaxy sample.

 \subsection{Model fits}

We perform a $\chi^2$ minimization in order to obtain constraints for the free parameters of the models.  
For a given free parameter $a$, we minimize the function 
\begin{equation}
     \chi^2(a)=\left[\bar{\omega}_{\mathrm{v}+}-\omega_{\mathrm{v}+}^\text{model}(a)\right]\mathcal{C}^{-1} \left[\bar{\omega}_{\mathrm{v}+}- \omega_{\mathrm{v}+}^\text{model}(a)\right]^{\top}.
\end{equation}
The best fit corresponds to the minimal value of $\chi^2$: $\chi^2_\text{model}$.
The confidence interval is determined by the requirement that
\begin{equation}
     \chi^2(x)<\chi^2_{\text{model}}+\delta\chi^2,
\end{equation}
with $\delta\chi^2$ depending on the number of free parameters and the confidence level. We report $68\%$ confidence levels as default, which sets $\delta\chi^2=1 $ for 1 free parameter. 
We also calculate the $\chi^2$ value for the null model: $\chi^2_0$, and the difference between both
\begin{equation}
     \Delta \chi^2 =\chi^2_0-\chi^2_\text{model}\geq 0.
\end{equation}

From $\chi^2_\text{model}$, $\chi^2_0$ and $\Delta \chi^2$, we calculate three p-values $p_\text{model}$, $p_0$ and $p_\Delta$. $p_\text{model }$ and $p_0$  evaluate the rejection or not of the model, and the null hypothesis, by the data. $p_\Delta $ gives an assessment on the significance of goodness of the best fit in comparison to the null hypothesis.

\section{Results}
\label{sec:4}
We restrict our measurements of the projected correlation (Eq. \ref{eq:estimfinal}) to scales $r_p>10\,h^{-1}$ Mpc. Smaller scales are affected by large error bars due to a paucity of galaxy-void pairs given the intrinsic low density of these environments \citep{Sutter_2014}. The intermediate-scale model (Eq. \ref{eq:int_scale_pred}) is fit at $r_p<1.5r_\mathrm{v}$, and the large-scale model (Eq. \ref{eq:large_scale_pred}) is adopted when $r_p>2r_\mathrm{v}$. We thus obtain a constraint for the free coefficients of each model: $A_I^Va_\mathrm{v}$ at intermediate scales (introduced in section \ref{sec:model_intscale}) and $A_Ib_\mathrm{v}(r_\mathrm{v})$  at large scales (introduced in section \ref{sec:model_largescale}). The bias of the voids depends on their radius~\citep{Hamaus_2013}, so we expect different values for each measurement.  

A summary of our results is presented in Table~\ref{tab:table_measurment} and discussed in the following subsections. In Appendix \ref{sec:appendixB} we show results for $\omega_{\mathrm{v}\times}$ to check its consistency with a null signal, which corresponds to a non-zero $p$-value. These measurements confirm that for $r_\mathrm{v}\in\,  20-30$ and $30-40 \, h^{-1}$ Mpc, there is no ${\times}$ correlation. 
For larger voids, we find a small $p$-value ($p_0<0.01$), which means that  $\omega_{\mathrm{v}\times}$ is not consistent with the null hypothesis. We  give some possible explanation of this measure in \ref{sec:appendixB3}, and will not exploit in detail the $+$ correlation for large voids since it might be affected by systematics.

Combining the three void sample measurements, we find no significant deviation from the null hypothesis for either intermediate- and large-scale regimes ($p_0=0.41$), the large-scale regime ($p_0=0.25$), or the intermediate-scale regime ($p_0=0.60$).

\begin{table*}
	\centering
	\begin{tabular}{lcccccccccr} 
		\hline
		$r_\mathrm{v}$ [$h^{-1}$ Mpc] &Regime & $\# $points & Scales fit [$h^{-1}$ Mpc] & $\chi_\text{model}^2$ &$p_{\text{model}}$ & $\chi_0^2$ &$p_0$& $\Delta \chi^2$& $p_{\Delta}$ & Bias factors \\
		\hline
		$[20,30]$ &Intermediate scales &6& 10-36 &10.0 & 0.075&11.3 &  0.046&1.3 & 0.93&$A_I^Va_\mathrm{v}=-123\pm 105$\\
		$[20,30]$ &Large scales &5&46-127 & 2.0&0.74 & 9.3 &0.054&7.3 & 0.12 & $A_Ib_\mathrm{v}=27.0\pm11.1$\\
		\hline
		$[30,40]$ &Intermediate scales &7& 10-46 &1.7 &0.94 & 4.1&0.63& 2.4& 0.88&$A_I^Va_\mathrm{v}=75\pm 45$\\
		$[30,40]$& Large scales &4&60-127 & 0.88 & 0.83& 5.8 &0.11&4.9 & 0.18&$A_Ib_\mathrm{v}=18.4\pm 8.2$\\
		\hline
		$[40,50]$ &Intermediate scales &8& 10-60 &2.2&0.99 &2.4 &0.94&0.2 & 1 &$A_I^Va_\mathrm{v}=-13\pm31$\\
		$[40,50]$ & Large scales &3&77-127 & 0.76&0.68 &0.82 & 0.66& 0.06 & 0.97& $A_Ib_\mathrm{v}=-3.9\pm16.1$\\
		\hline
	\end{tabular}
	
	\caption{Results of intrinsic alignment amplitude fits from $\omega_{\mathrm{v}+}$ measurements in both regimes and for the different void radius ranges. Columns 5 and 6 indicate the quality of the best fit via the chi-squares and p-values. Columns 7 and 8 show the corresponding values for the null hypothesis (no signal). Column 9 provides $p_{\Delta}$, the probability for rejection of the null hypothesis, assuming the model is valid.
	\label{tab:table_measurment}}
\end{table*}

\subsection{Small voids}

\begin{figure}
    \centering
    \includegraphics[scale=0.58]{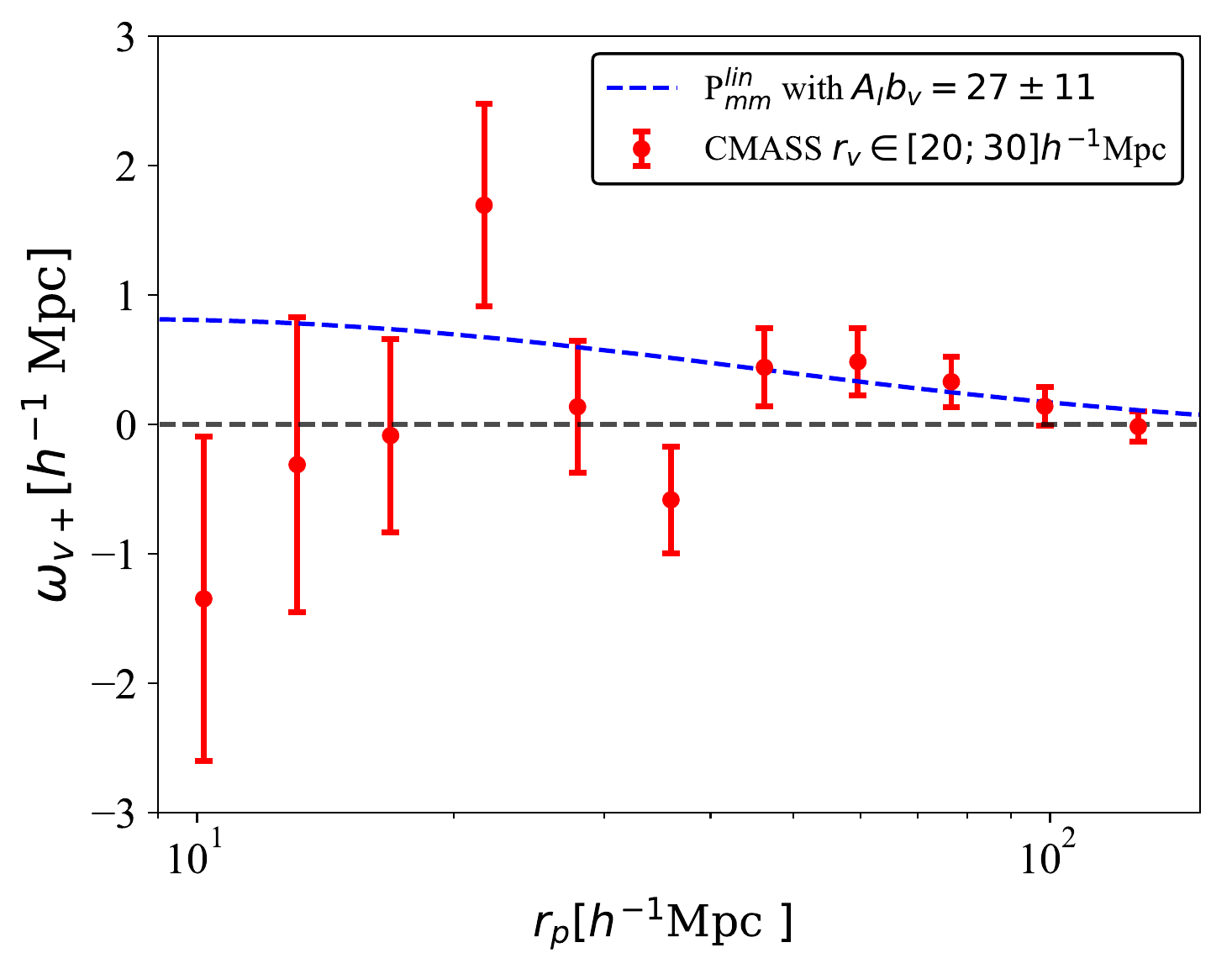}
    \caption{$\omega_{\mathrm{v}+}$ with $r_\mathrm{v}=[20,30]\,h^{-1}$ Mpc, and the large-scale prediction with $A_Ib_\mathrm{v}=27.0\pm11.1$. The intermediate-scale model is rejected, and we did not include it in the plot. 
    }
    \label{fig:paternsGLIA1}
\end{figure}

The measurement of $\omega_{\mathrm{v}+}$ for this sample of voids is shown in Fig.~\ref{fig:paternsGLIA1}, along with the resulting large-scale model fit, applied to scales $r_p>46\,h^{-1}$ Mpc (5 points). We do not show the best fit with the intermediate-scale model, because it has been rejected by the data at $1.8 \sigma$ : $p_\text{model}=0.075$. The null hypothesis is rejected at 2$\sigma$ for both fit ranges ($p_0\sim$ 0.05 for both regimes).\\
For the large-scale model, we find $A_Ib_\mathrm{v}(r_\mathrm{v})=27.0\pm11.1$, with $p_\text{model}=0.74$.  
Here, the large-scale model describes the measurements well. Nonetheless, $p_\Delta=0.12$, which means that the significance for a detection of the alignment is lower than 1.5$\sigma$.

\subsection{Intermediate voids}
\begin{figure}
    \centering
    \includegraphics[scale=0.58]{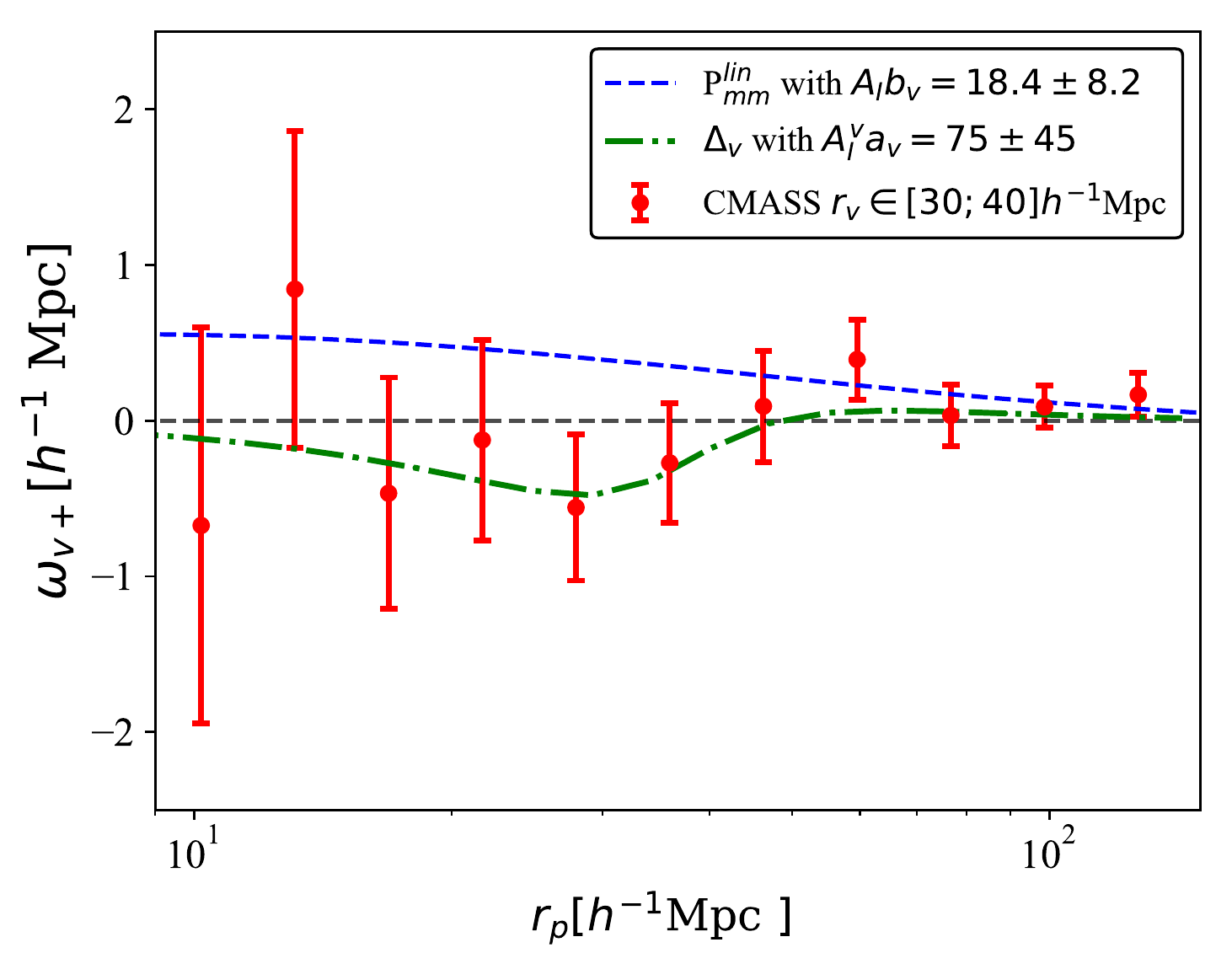}
    \caption{$\omega_{\mathrm{v}+}$ with $r_\mathrm{v} \in [30;40]\,h^{-1}$ Mpc, the intermediate-scale prediction with $A_I^Va_\mathrm{v}=75\pm45$, and the large-scale one with $A_Ib_\mathrm{v}=18.4\pm8.2$.
    }
    \label{fig:paternsGLIA2}
\end{figure}

We use the intermediate-scale model for $10h^{-1}\mathrm{Mpc}<r_p<46h^{-1}\mathrm{Mpc}$ (7 points) and the large-scale one for $r_p>60h^{-1}\mathrm{Mpc}$ (4 points) for the intermediate-size voids. The measurements and both models are shown in Fig.~\ref{fig:paternsGLIA2}. The null hypothesis is rejected at 1.5$\sigma$ on large scales ($p_0=0.11$), but it is not rejected at small scales ($p_0=0.63$).  \\
In the intermediate-scale regime, we find $A_I^Va_\mathrm{v}=75\pm45$, with $p_\text{model}=0.94$ confirming a good fit. Nonetheless, since $p_\Delta=0.88$, our model is not significantly better than the null one.
In the large-scale regime, we find $A_Ib_\mathrm{v}(r_\mathrm{v})=18.4\pm8.2$, with $p_\text{model}=0.83$. The detection of the alignment is not statistically significant, since $p_\Delta=0.18$.

\subsection{Large voids}

\begin{figure}
    \centering
    \includegraphics[scale=0.58]{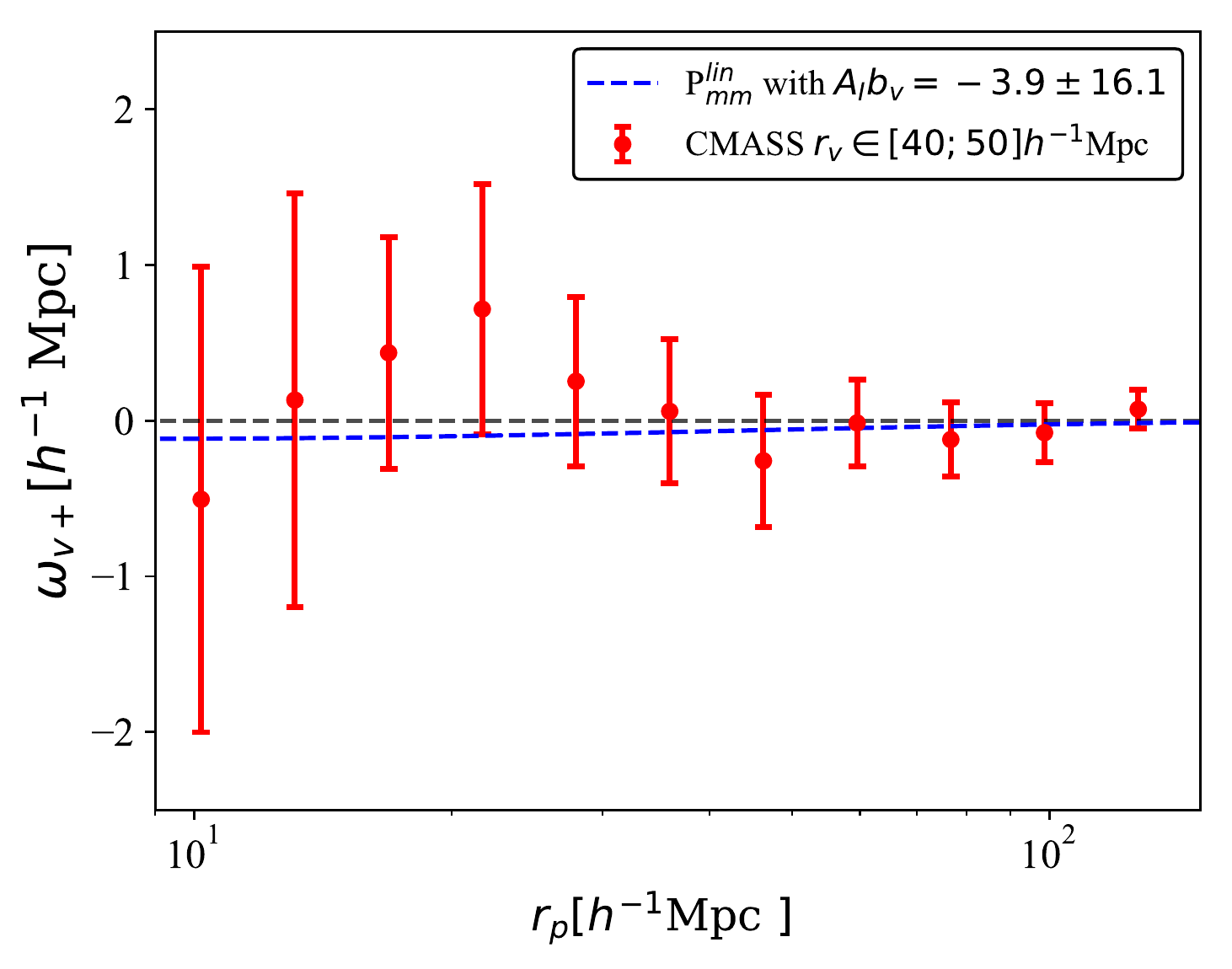}
    \caption{$\omega_{\mathrm{v}+}$ with $r_\mathrm{v} \in [40;50]h^{-1}\mathrm{Mpc}$, and the large-scale prediction with $A_Ib_\mathrm{v}=-3.9\pm 16.1$. The signal at intermediate scales is too noisy to have a detection of alignments in the intermediate-scale regime. 
    }
    \label{fig:paternsGLIA3}
\end{figure}

We use the intermediate-scale model for $r_p<60$ $h^{-1}$Mpc (8 points) and the large-scale one for $r_p>77$ $h^{-1}$Mpc (3 points). The measurement is shown in Figure \ref{fig:paternsGLIA3}, with the resulting large-scale model overlaid. Here again, we do not show the intermediate-scale one, because there is no distinction between the best fit, and the null one: $p_\Delta=1$. The null hypothesis is not excluded at both scales: $p_0=0.94$ for intermediate scales and $p_0=0.66$ for large scales.\\
On large scales, we find $A_Ib_\mathrm{v}(r_\mathrm{v})=-3.9\pm16.1$, with $p_\text{model}=0.68$.  There is almost no distinction between the best fit and the null hypothesis, since $p_\Delta=0.97$. 

\section{Discussion}
\label{sec:discussion}
Following our modelling, we expect to detect two distinct regimes:
\begin{itemize}
    \item at intermediate scales, due to the proximity of the void, the correlation should be negative, i.e. the galaxies align tangentially; 
    \item at larger scales, the alignment depends on the value of the void bias; for positive bias, the galaxies align radially (pattern of figure \ref{fig:paternsGLIA}); for negative bias, they align tangentially.
\end{itemize}
However, these effects, especially for the intermediate scales, are supposed to be weak and therefore difficult to detect; especially since we have a limited sample of only a few thousand voids. 

For voids with radii between 20 and 30$h^{-1}\mathrm{Mpc}$, we measured alignments that were compatible with no signal at 2$\sigma$. For these voids we also do not measure any signature of the void contribution at intermediate scales. This may be due to the tendency of these voids to contain more tracers at their boundary. At large scales, we find an IA amplitude of $A_Ib_\mathrm{v}(25 \,h^{-1}\text{Mpc})\sim 27 \pm 11.1$. This is an acceptable order of magnitude, since we expect $A_I \sim 5$ as for LOWZ and CMASS (\citealt{Singh,Singh2021FP}), which would mean that we have a void bias around 2-6, (in order of magnitude similar to  \citealt{Clampitt_2016}). The analysis of the $\times$ correlation between shapes and voids in appendix \ref{sec:appendixB1} confirms that this is compatible with null ($p_0 = 0.96$), as expected.
 
For voids with intermediate sizes, the null hypothesis at large scales is almost excluded by the p-value test ($p_0=0.11$), but it is not excluded at intermediate scales ($p_0=0.63$). Given the noise, and the fact that even the model only deviates from 0 at a small interval at $r_p$ close to $r_\mathrm{v}$, it is not surprising to obtain a high $p$-value for the null hypothesis, and we were not able to distinguish statistically the null model from the best fit one: $p_\Delta=0.88$. 
At large scales, the model agrees relatively well with the measurement, with a measured IA coefficient $A_I b_\mathrm{v}(35 \,h^{-1}\text{Mpc})\sim 18.4\pm 8.2$.  
We find here again a physically acceptable order of magnitude. Since $A_I$ is the same as in the first measurement (we are working with the same galaxy shape sample), the void bias for $r_\mathrm{v}\sim 35 \,h^{-1}$ Mpc appears to be smaller than the one for $r_\mathrm{v}\sim 25 \,h^{-1}$ Mpc. A trend of decreasing bias with increasing void radius has already been suggested by \citet{Hamaus_2013,Chan_2014,separateunivsersevoidbias,Chan_2020} from simulations. 
The analysis of the $\times$ correlation between shapes and voids in appendix \ref{sec:appendixB}  confirms again that this is compatible with null as expected, with $p_0= 0.17$.

Finally, the intrinsic alignments of CMASS galaxies around large voids are compatible with null ($p_0=0.66$ for large scales and $p_0=0.94$ for intermediate scales). Moreover, the $\times$ correlation between shapes and voids, presented in appendix \ref{sec:appendixB3}, is not compatible with 0 ($p_0=9\times 10^{-4}$), which could indicate that the measurements are not reliable. This may be due to the smaller number of large voids, and their more complex sub-structure, which could break the assumption of spherical symmetry. It is also possible that at large scales the null measurement of the $+$ correlation is due to a void bias close to 0, which is typical for compensated voids \citep{Hamaus_2013}. 

\section{Conclusion and perspectives}
\label{sec:5}

We have studied intrinsic alignments of galaxies around cosmic voids, in SDSS-III BOSS CMASS galaxies. Using this sample, we have investigated the shape-position correlation from 10 to 140 $h^{-1}$ Mpc scales, binning voids by size.   

Our model suggests the existence of two regimes: for scales equivalent to the void radius, the correlation should be negative, which means that galaxies align tangentially. At large scale, for positive void bias, the correlation should be positive, which implies a radial alignment for galaxies (negative bias implies a tangential alignment). The intermediate-scale regime with the negative correlation is more difficult to observe, because the available number of galaxies close to the void centre is small. However, it may be the most interesting one in order to extract physical information and constraints, since it directly depends on the mass distribution inside voids.

We find that the large-scale model fits well data for void radius ranges: 20-30 $h^{-1}$ Mpc and 30-40 $h^{-1}$ Mpc, giving constraints for the void bias. For the void radius range 40-50 $h^{-1}$ Mpc, the measurement is consistent with 0 at any scale. For the large-scale part, it may be the consequence of a vanishing void bias. Furthermore, we find a significant null model deviation for the $\times$ correlation, which implies to be careful with this last measurement.

For the intermediate-scale regime, as expected, the noise is the main limit for the detection of a correlation. Our intermediate-scale model seems  ineffective to describe the measurement for 20-30 $h^{-1}$ Mpc and 40-50 $h^{-1}$ Mpc void radius ranges, because the model is rejected by data for smaller voids, and equivalent compatible with null for larger voids. Keeping only voids with radius between 30 and 40 $h^{-1}$ Mpc, we achieve a tentative detection of a negative correlation in agreement with the predictions of our intermediate-scale model. Smaller voids are more clustered than bigger ones, which may lead to a strong positive correlation coming from high-density environments. In such a case, the contribution from the voids would be hard to detect. Large voids are not very numerous, which combined with an important proportion of voids being elongated and featuring sub-structure, may contradict the spherical symmetry assumption of our model and may lead to biased measurements. Furthermore, voids with radius between 30-40 $h^{-1}$ Mpc are the most abundant, which helps to increase the signal-to-noise ratio and reduce potential systematics coming from small-number statistics.

The number of voids is supposed to drastically increase with future surveys by about two orders of magnitude~\citep{pisani2019cosmic}, which will necessarily lead to a substantial reduction of the noise due to the noise contribution to the covariance scaling directly with the inverse of the number density for both voids and galaxy shapes.

Thus, future surveys will allow us to confirm or rule out the tentative detection of a negative alignment correlation at intermediate scale for voids of all sizes.  With a noise reduction, it would also be possible to make a joint analysis with other correlations to remove the degeneracies of the coefficients, and extract physical constraints. 

In anticipation, we plan to investigate the validity of our model with numerical simulations. This could open the door for new applications of intrinsic alignments and for calculating their contamination to void lensing studies with photometric or spectroscopic
galaxy samples \citep[e.g.,][]{lensingvoids,NH5,NH6}.

\section*{Acknowledgements}
 We would like to thank C. Vedder, I.R. van Gemeren and R. Nederstigt for their help during our group meetings, M. Noble, T. Knibiehly and L.P. Chaintron for their help with algorithm optimization. 
 We are grateful to Harry Johnston for helping us with the $K$-means implementation to build random catalogues. We thank Rachel Mandelbaum for sharing the catalogue of galaxy shapes with us. 
 Also we would like to thank  N. Kaiser, for his encouragement at the beginning of the study. 
 NH is supported by the Excellence Cluster ORIGINS, which is funded by the Deutsche Forschungsgemeinschaft (DFG, German Research Foundation) under Germany's Excellence Strategy -- EXC-2094 -- 390783311. This work is part of the Delta ITP consortium, a program of the Netherlands Organisation for Scientific Research (NWO) that is funded by the Dutch Ministry of Education, Culture and Science (OCW), project number 24.001.027.
 
 Funding for SDSS-III has been provided by the Alfred P. Sloan Foundation, the Participating Institutions, the National Science Foundation, and the U.S. Department of Energy Office of Science. The SDSS-III web site is http://www.sdss3.org/. SDSS-III is managed by the Astrophysical Research Consortium for the Participating Institutions of the SDSS-III Collaboration including the University of Arizona, the Brazilian Participation Group, Brookhaven National Laboratory, Carnegie Mellon University, University of Florida, the French Participation Group, the German Participation Group, Harvard University, the Instituto de Astrofisica de Canarias, the Michigan State/Notre Dame/JINA Participation Group, Johns Hopkins University, Lawrence Berkeley National Laboratory, Max Planck Institute for Astrophysics, Max Planck Institute for Extraterrestrial Physics, New Mexico State University, New York University, Ohio State University, Pennsylvania State University, University of Portsmouth, Princeton University, the Spanish Participation Group, University of Tokyo, University of Utah, Vanderbilt University, University of Virginia, University of Washington, and Yale University.
 
\section*{Data availability}

No new data were generated or analysed in support of this research. The shape catalogue was presented in \cite{Singh2021FP} and the void catalogue, in \cite{hamaus2020precision}.




\bibliographystyle{mnras}
\bibliography{bibli}




\appendix
\section{Impact of density-weighting of galaxy shapes} 
\label{sec:appendixA}

The ${\gamma}^I$ field is not uniformly sampled by observations. Rather, we sample it at the locations of galaxies. Thus in reality, we observe
\begin{equation}
    \Tilde{\gamma}^I=(1+b_\mathrm{g}\delta)\gamma^I.
\end{equation}
 Using Fourier transformation on $\Tilde{\gamma}^I$, we  obtain an integral expression, 
\begin{equation}
\begin{split}
    &\Tilde{\gamma}^I({\bf k}~,~z)=-\frac{C_1 \bar{\rho}_\text{m}}{D(z)}\iiint d^3k_1\\
    &\left\{ \left(k_{2x}^2-k_{2y}^2~,~2k_{2x}k_{2y}\right) \delta({\bf k_2},z) \left(\delta^{(3)}({\bf k_1})+b_\mathrm{g}\delta({\bf k_1}~,~z)\right)\right\},
    \end{split}
\end{equation}
with $\bf k_2=k-k_1$.
Here, we directly see that in the Fourier space, 
\begin{equation}
    \langle \delta_\mathrm{v} \tilde{\gamma_I}\rangle=\langle \delta_\mathrm{v} \gamma_I\rangle+\mathcal{O}(\delta_\mathrm{v}\delta_\mathrm{m}^2)
    \end{equation}\
Thus, if we restrict our study to linear order, we can omit the correction to $\gamma_I$.

\section{$\times$ correlation}
\label{sec:appendixB}
We present here the measurements of the $\times$ correlation, which we expect to be consistent with a null signal. Otherwise, it could be a possible indication that systematics affect the measurement. We estimate the $\times$ correlation using Eq.~(\ref{eq:estimator}), with $S_\times$ instead of $S_+$. We compute the $p$-value for each measurement (corresponding to different void radii), and summarize the results in Table \ref{tab:cross}.
\begin{table}
	\centering
	
	\begin{tabular}{lr} 
		\hline
		$r_\mathrm{v}$ [$h^{-1}$ Mpc] & $p_0$ \\
		\hline
		[20-30] &$0.96$\\
		\hline
		[30-40] &$0.17$\\
		\hline
		[40-50] &$0.0009$\\
		\hline
	\end{tabular}
	\caption{Table of the p-values of the null hypothesis for the $\times$ correlation.  \label{tab:cross}}
\end{table}

\subsection{Void radius in 20-30 $h^{-1}$ Mpc range}
\label{sec:appendixB1}
\begin{figure}
    \centering
    \includegraphics[scale=0.4]{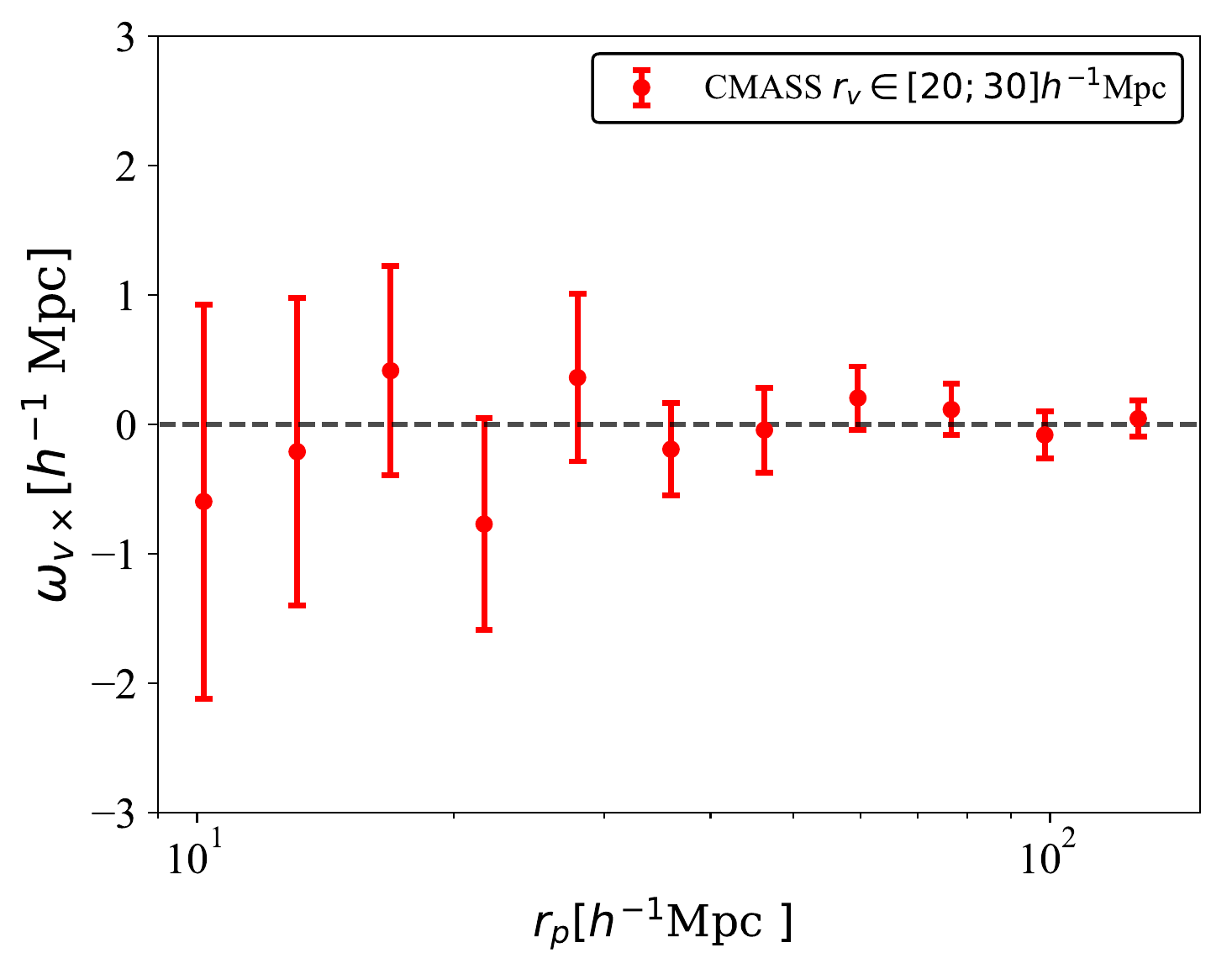}
    \caption{$\times$ correlation for voids with $r_\mathrm{v}\in$ 20-30 $h^{-1}$ Mpc.}
    \label{fig:cross_20_30}
\end{figure}

The measurement is presented in figure \ref{fig:cross_20_30}. For small voids, we find a $p$-value compatible with a null signal: $p_0=0.96$.

\subsection{Void radius in 30-40 $h^{-1}$ Mpc range}
\label{sec:appendixB2}
\begin{figure}
    \centering
    \includegraphics[scale=0.4]{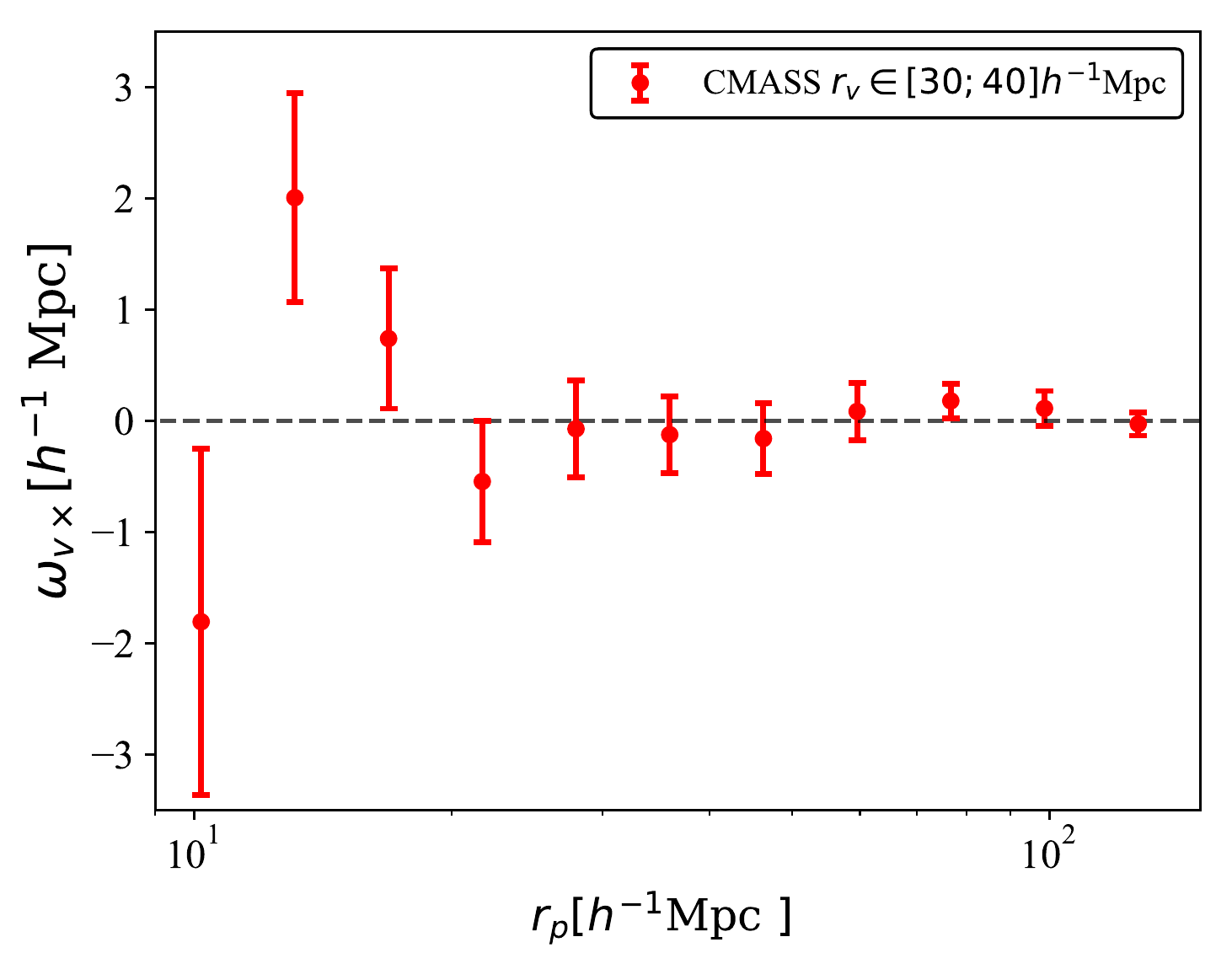}
    \caption{$\times$  correlation for voids with $r_\mathrm{v}\in$ 30-40 $h^{-1}$ Mpc.}
    \label{fig:cross_30_40}
\end{figure}

The measurement is presented in figure \ref{fig:cross_30_40}. We find a $p$-value quite compatible with a null signal: $p_0=0.16$, even if its value is lower than for smaller voids, mainly because of the positive trend at $r_p\sim 10h^{-1}\mathrm{Mpc}$.

\subsection{Void radius in 40-50 $h^{-1}$ Mpc range}
\label{sec:appendixB3}
\begin{figure}
    \centering
    \includegraphics[scale=0.4]{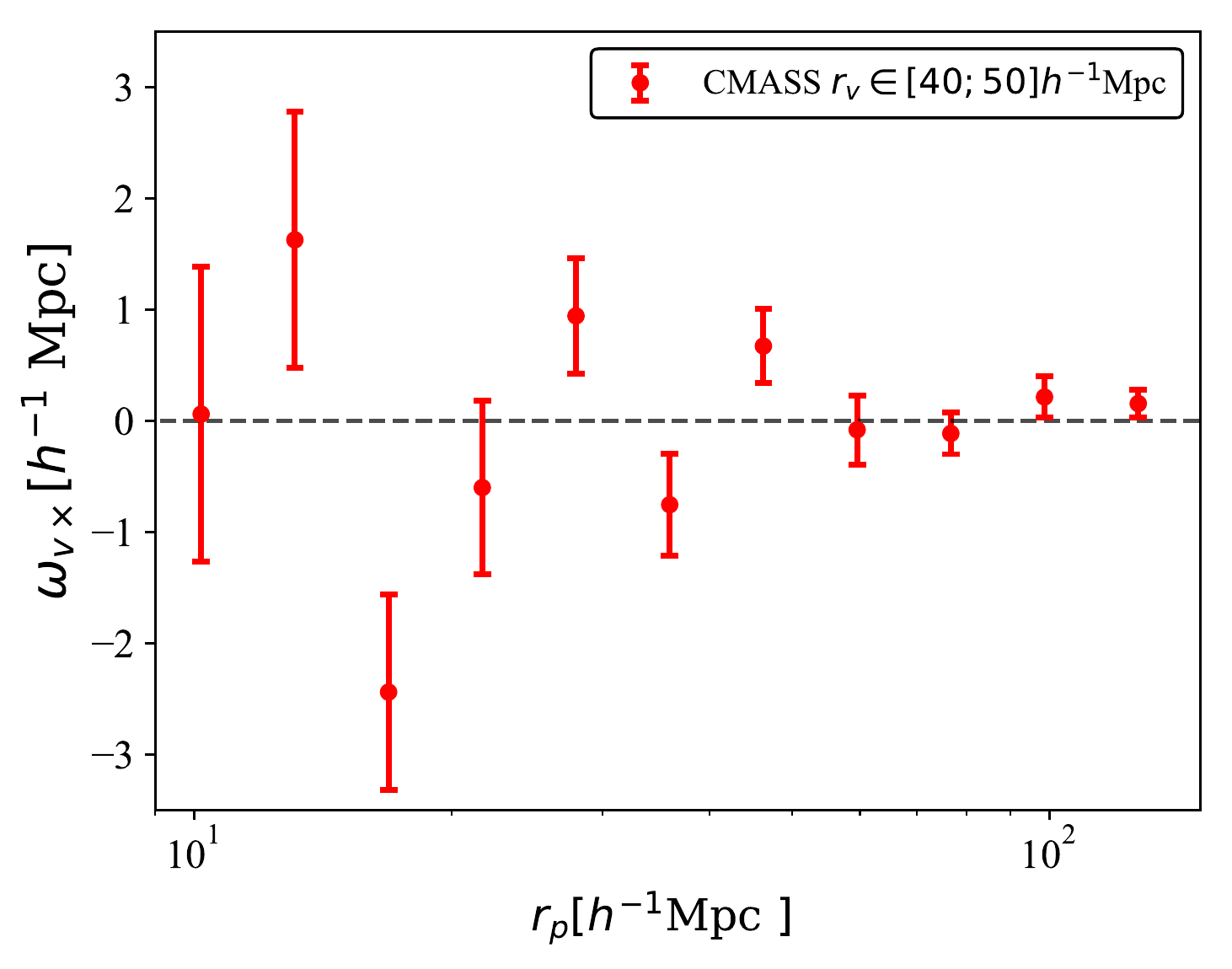}
    \caption{$\times$ correlation for voids with $r_\mathrm{v}\in$ 40-50 $h^{-1}$ Mpc.}
    \label{fig:cross_40_50}
\end{figure}

The measurement is presented in figure \ref{fig:cross_40_50}. For large voids, we find a $p$-value that suggests the rejection of the null hypothesis: $p_0=0.0009$. The measurements for these voids are not exploited in detail in the analysis presented in the main body of the manuscript, since they may be subject to systematics. It is also possible that we detect a non-vanishing signal due to important sub-structure inside voids, and the void sample could not be large enough to average out their contribution. Alternatively, the error bars could be underestimated.  
\label{lastpage}
\end{document}